\documentclass[aps,twocolumn,prl,showpacs,amsmath,amssymb,10pt]{revtex4-2}
\usepackage{graphicx}
\usepackage{scrextend}
\usepackage{manfnt,pifont}
\usepackage{hyperref}
\usepackage{url}
\usepackage{lipsum}
\usepackage[font=small]{subcaption}
\usepackage[font=small]{caption}

\usepackage[normalem]{ulem}

\usepackage{url}
\usepackage[outline]{contour}
\usepackage{textgreek}

\usepackage{tikz}
\usetikzlibrary{shapes}

\usepackage{CJK}
\usepackage{enumitem}
\setlist[itemize]{leftmargin=*}

\newcommand\wh\widehat

\makeatletter
\DeclareFontFamily{OMX}{MnSymbolE}{}
\DeclareSymbolFont{MnLargeSymbols}{OMX}{MnSymbolE}{m}{n}
\SetSymbolFont{MnLargeSymbols}{bold}{OMX}{MnSymbolE}{b}{n}
\DeclareFontShape{OMX}{MnSymbolE}{m}{n}{
    <-6>  MnSymbolE5
   <6-7>  MnSymbolE6
   <7-8>  MnSymbolE7
   <8-9>  MnSymbolE8
   <9-10> MnSymbolE9
  <10-12> MnSymbolE10
  <12->   MnSymbolE12
}{}
\DeclareFontShape{OMX}{MnSymbolE}{b}{n}{
    <-6>  MnSymbolE-Bold5
   <6-7>  MnSymbolE-Bold6
   <7-8>  MnSymbolE-Bold7
   <8-9>  MnSymbolE-Bold8
   <9-10> MnSymbolE-Bold9
  <10-12> MnSymbolE-Bold10
  <12->   MnSymbolE-Bold12
}{}

\let\llangle\@undefined
\let\rrangle\@undefined
\DeclareMathDelimiter{\llangle}{\mathopen}%
                     {MnLargeSymbols}{'164}{MnLargeSymbols}{'164}
\DeclareMathDelimiter{\rrangle}{\mathclose}%
                     {MnLargeSymbols}{'171}{MnLargeSymbols}{'171}
\makeatother

\usepackage{eso-pic}
\newcommand{\reportnum}[2]{
  \AddToShipoutPictureBG*{%
    \AtPageUpperLeft{%
      \hspace{0.75\paperwidth}%
      \raisebox{#1\baselineskip}{%
        \makebox[0pt][l]{\textnormal{#2}}
  }}}%
}

\begin{document}

\reportnum{-3}{USTC-ICTS/PCFT-25-14}

\title{Giant Graviton Correlators as Defect Systems}

\author{Junding Chen$^{a,b}$}
\author{Yunfeng Jiang$^{c,d}$}
\email{jinagyf2008@seu.edu.cn}
\author{Xinan Zhou$^{b,d}$}
\email{xinan.zhou@ucas.ac.cn}
\affiliation{$^a$School of Physical Sciences, University of Chinese Academy of Sciences, No.19A Yuquan Road, Beijing 100049, China}
\affiliation{$^b$Kavli Institute for Theoretical Sciences, University of Chinese Academy of Sciences, Beijing 100190, China}
\affiliation{$^c$School of physics \& Shing-Tung Yau Center, Southeast University,
Nanjing 211189, P. R. China}
\affiliation{$^{d}$Peng Huanwu Center for Fundamental Theory, Hefei, Anhui 230026, China}

\begin{abstract}
We consider correlation functions of two maximal giant gravitons and two light $\frac{1}{2}$-BPS operators in 4d $\mathcal{N}=4$ SYM. Viewed as two-point correlators in the presence of a zero dimensional defect, they can be completely fixed at strong coupling using analytic bootstrap techniques. We determine all infinitely such correlators for arbitrary light $\frac{1}{2}$-BPS operators and find that the result can be repackaged into a simple generating function thanks to a hidden higher dimensional symmetry. We also find evidence that the same symmetry holds at weak coupling for loop correction integrands. 

\end{abstract}

\maketitle

\noindent{\bf Introduction.} Defects are important extended objects, the studies of which have a wide range of motivations, both experimental (impurities, boundaries, domain walls, etc) and theoretical (symmetry generators, D-branes, etc). Adding defects enriches theories by introducing new observables and data. On the one hand, defects usually admit localized degrees of freedom  which require new physical parameters to characterize. On the other, defects break symmetries and thereby allow new observables to arise which are otherwise prohibited. These interesting features make defects an exciting arena where we can go beyond just theories in infinite empty space.

In the canonical testing ground of 4d $\mathcal{N}=4$ SYM there are already many different types of defects. Categorized by dimensions $p$, the best known examples with $p=1$ are Wilson and 't Hooft lines. The case of $p=2$ corresponds to surface defects \cite{Gukov:2006jk,Buchbinder:2007ar,Drukker:2008wr}. For $p=3$, we can have interfaces and boundaries \cite{DeWolfe:2001pq,Bak:2003jk,Gaiotto:2008sa}. Objects with these three values of $p$ are what we usually perceive as defects. However, this catalogue can be further extended by including defects with dimensions $p=-1,0$. These seemingly exotic values become easy to understand via holography where an extra dimension is gained by moving into AdS. In particular,  $\mathcal{N}=4$ SYM on $\mathbb{RP}^4$ can be viewed as an example of a $p=-1$ dimensional defect. This is because the holographic dual is a $\mathbb{Z}_2$ quotient of AdS$_5\times$S$^5$ which has a fixed locus sitting at a point in AdS$_5$ \cite{Wang:2020jgh,Caetano:2022mus}. An example of $p=0$ defect is provided by giant gravitons \cite{McGreevy:2000cw,Balasubramanian:2001nh,Hashimoto:2000zp}. This follows from that giant gravitons are heavy states travelling along one dimensional geodesics in AdS. 
Defects therefore provide a unifying perspective despite very different physical interpretations of these systems. This defect viewpoint, which may not seem much at the moment, turns out to be very useful, as will be demonstrated in this paper. 

We focus on here the last example of giant gravitons. The other unconventional case of defects with $p=-1$ has been considered in \cite{Zhou:2024ekb}. However, under comparison the giant graviton correlators are especially interesting for two reasons. First, usually correlators of light operators are considered and studies involving heavy operators are far fewer. Let us consider correlators of two giant graviton operators with other light operators. Most results concern the situation where there is only one light operator \cite{Bissi:2011dc,Caputa:2012yj,Jiang:2019xdz,Jiang:2019zig,Holguin:2022zii}. This is somewhat less exciting since symmetries determine three-point functions up to  overall coefficients. For two light operators, the functional form of correlators is  unconstrained but not much is known. At weak coupling, such four-point functions have been computed up to three loops \cite{Jiang:2019xdz,Jiang:2019zig,Jiang:2023uut}. But at strong coupling, there is no result except for a prediction for the integrated correlator when both light operators are in the stress tensor multiplet \cite{Brown:2024tru}. It appears that there is still a lot of progress to be made. Second, this setup is a rare situation where  correlators of local operators admit dual interpretations. As we mentioned, these four-point functions can also be viewed as two-point functions of light operators in the presence of a zero dimensional defect. This duality allows us to leverage the well established results of four-point function kinematics \cite{Eden:2000bk,Nirschl:2004pa} and at the same time apply defect system technologies \cite{Billo:2016cpy}. 

The results of this paper can be summarized as follows. We study four-point functions of two maximal giant gravitons and two light $\frac{1}{2}$-BPS operators in the large $N$ limit. At strong coupling, AdS/CFT gives rise to a defect picture for these correlators where two supergravitons weakly interact with the geodesic line traversed by the heavy giant gravitons. With minimal input from a supergravity analysis, we apply the analytic bootstrap techniques of \cite{Rastelli:2016nze,Rastelli:2017udc,Gimenez-Grau:2023fcy,Chen:2023yvw,Zhou:2024ekb} to compute these correlators. We find that supersymmetry fixes all infinitely many correlators at tree level where the light $\frac{1}{2}$-BPS operators can have arbitrary dimensions. Quite remarkably, all correlators can be repackaged into a simple generating function thanks to a hidden higher dimensional symmetry. This hidden symmetry appears to also naturally extend to the weak coupling regime.

\vspace{0.5cm}

\noindent{\bf Kinematics.} Maximal giant gravitons are $\frac{1}{2}$-BPS operators built from the six scalars $\Phi^{I=1,\ldots,6}$ in $\mathcal{N}=4$ SYM
\begin{equation}
\nonumber \mathcal{D}(x,t)=\det(\Phi^I(x)t^I)\;,   
\end{equation}
where $t^I$ is a null R-symmetry polarization vector with $t\cdot t=0$. For $SU(N)$ gauge group, giant gravitons have protected dimension $\Delta=N$ and R-symmetry representation $[0,N,0]$. The light $\frac{1}{2}$-BPS operators, or the  supergravitons, are defined as 
\begin{equation}
\nonumber    \mathcal{O}_k(x,t)=\mathcal{N}_k\, {\rm tr} (\Phi^I(x)t^I)^k\;,
\end{equation}
which have $\Delta=k$ and transform in the $[0,k,0]$ representation. Here $\mathcal{N}_k$ is chosen such that the two-point function is normalized to one. We are interested in the four-point functions
\begin{equation}
\nonumber    F_{k_1k_2}(x_i,t_i)=\langle \mathcal{O}_{k_1}(x_1,t_1)\mathcal{O}_{k_2}(x_2,t_2)\mathcal{D}(x_3,t_3)\mathcal{D}(x_4,t_4) \rangle\;,
\end{equation}
and now we discuss their kinematics from two complementary perspectives. 
\vspace{0.5cm}

\noindent{\it -- As four-point functions.} $F_{k_1k_2}$ is constrained by superconformal symmetry to take the form \cite{Eden:2000bk,Nirschl:2004pa}
\begin{equation}
\nonumber    F_{k_1k_2}(x_i,t_i)=F_{k_1k_2}^{\rm free}(x_i,t_i)+t_{12}^2t_{34}^2x_{13}^4x_{24}^4 R\, K_{k_1k_2}(x_i,t_i)\;,
\end{equation}
where $t_{ij}=t_i\cdot t_j$, $x_{ij}=x_i-x_j$ and $F_{k_1k_2}^{\rm free}$ is the correlator in free theory. Let us define the cross ratios 
\begin{equation}
\begin{split}
\nonumber {}&U=\frac{x_{12}^2x_{34}^2}{x_{13}^2x_{24}^2}=z\bar{z}\;,\quad V=\frac{x_{14}^2x_{23}^2}{x_{13}^2x_{24}^2}=(1-z)(1-\bar{z})\;,\\ 
{}& \sigma=\frac{t_{13}t_{24}}{t_{12}t_{34}}=\alpha\bar{\alpha}\;,\quad \tau=\frac{t_{14}t_{23}}{t_{12}t_{34}}=(1-\alpha)(1-\bar{\alpha})\;.
\end{split}
\end{equation}
Then $R$ has the form
\begin{equation}
R=(1-z\alpha)(1-z\bar{\alpha})(1-\bar{z}\alpha)(1-\bar{z}\bar{\alpha})\;,
\end{equation}
and $K_{k_1k_2}(x_i,t_i)$ is a reduced four-point correlator with shifted R-symmetry charges $\{k_1-2,k_2-2,N-2,N-2\}$ and conformal dimensions $\{k_1+2,k_2+2,N+2,N+2\}$. 
\vspace{0.5cm}

\noindent{\it -- As defect two-point functions.} At the kinematic level, we can always view $F_{k_1k_2}$ as the two-point function of $\mathcal{O}_{k_1}$ and $\mathcal{O}_{k_2}$ in the presence of a zero dimensional defect created by the pair of $\mathcal{D}$'s. Concretely, we divide $F_{k_1k_2}$ by the $\langle\mathcal{D}\mathcal{D}\rangle$ two-point function and define
\begin{equation}
\nonumber G_{k_1k_2}\equiv \llangle \mathcal{O}_{k_1}(x_1,t_1)\mathcal{O}_{k_2}(x_2,t_2) \rrangle = \frac{F_{k_1k_2}(x_i,t_i)}{\langle \mathcal{D}(x_3,t_3)\mathcal{D}(x_4,t_4) \rangle}\;,
\end{equation}
which leads to 
\begin{equation}
\begin{split}\label{defGandH}
    G_{k_1k_2}={}&\frac{x_{34}^{2N}}{t_{34}^N}\left(F_{k_1k_2}^{\rm free}+t_{12}^2t_{34}^2x_{13}^4x_{24}^4 R\, K_{k_1k_2}\right)\\
    ={}&G_{k_1k_2,{\rm free}}+\frac{t_{12}^2x_{13}^4x_{24}^4}{x_{34}^4}R\, H_{k_1k_2}\;.
\end{split}
\end{equation}
Here we have similarly defined a reduced defect two-point function $H_{k_1k_2}$ which has shifted R-symmetry charges $\{k_1-2,k_2-2\}$ and conformal dimensions $\{k_1+2,k_2+2\}$. To proceed, it is convenient to manifest conformal symmetry by introducing embedding space vectors $P\in \mathbb{R}^{5,1}$ which are related to $x\in\mathbb{R}^4$ via
\begin{equation}
    P=\left(\frac{1+x^2}{2},\frac{1-x^2}{2},\vec{x}\right)\;,
\end{equation}
with signature $(-,+,+,\ldots)$. Defects can be characterized by the symmetries which they break. This is further achieved by introducing projectors. For the spacetime part, we define
\begin{equation}\label{defN}
    \mathbb{N}_{AB}=\frac{P_{3,A}P_{4,B}+P_{4,A}P_{3,B}}{P_3\cdot P_4}\;,
\end{equation}
which breaks conformal group $SO(5,1)\times SO(1,1)$ into $SO(4)$. We also define an R-symmetry projector
\begin{equation}\label{defM}
\mathbb{M}_{IJ}=\delta_{IJ}-\frac{t_{3,I}t_{4,J}+t_{4,I}t_{3,J}}{t_3\cdot t_4}\;,
\end{equation}
which reduces $SO(6)_R$ to $SO(4)\times SO(2)$. Note that $\mathbb{N}$ and $\mathbb{M}$ are similar but different. While $\mathbb{N}$ projects vectors to the plane spanned by $P_3$ and $P_4$, $\mathbb{M}$ projects vectors to the space orthogonal to $t_3$ and $t_4$. Clearly, symmetries determine one-point functions up to overall coefficients
\begin{eqnarray}
\nonumber \llangle \mathcal{O}_p(x_i,t_i)\rrangle\sim \frac{2^{-\frac{p}{2}}(t_i\cdot \mathbb{M}\cdot t_i)^{\frac{p}{2}}}{(P_i\cdot \mathbb{N}\cdot P_i)^{\frac{p}{2}}}=\left(\frac{t_{i3}t_{i4}x_{34}^2}{t_{34}x_{i3}^2x_{i4}^2}\right)^{\frac{p}{2}},\; p \text{ even}\;,
\end{eqnarray}
which also stems from the same statement for three-point functions in defect-free CFTs.  Factoring out one-point functions, the defect two-point correlator can be expressed in terms of defect cross ratios which are formed from ratios of $P_i\cdot P_j$ and $P_i\cdot \mathbb{N} \cdot P_j$ with $i,j=1,2$ invariant under $P_i\to\lambda_i P_i$ rescalings (similarly for the R-symmetry part). These are obviously related to the $U$, $V$, $\sigma$, $\tau$ via straightforward albeit nontrivial relations.

\vspace{0.5cm}

\noindent{\bf Supergravity description.} Giant gravitons are dual to D3-branes which wrap an $\mathbb{R}\times$S$^3$ subspace of  AdS$_5\times$S$^5$ where S$^3\subset $S$^5$. The dynamics is encoded in the action 
\begin{equation}
\label{action}
    S_{\rm D3}=-\frac{N}{2\pi^2}\int_{\mathbb{R}\times \text{S}^3} \left(\sqrt{\operatorname{det} g^{\rm P.B.}_{\alpha\beta}}-C^{\rm P.B.}_4\right)\;,
\end{equation}
where P.B. denotes the pull-back to the brane. Without loss of generality, we insert the giant gravitons at $x_3=0$ and $x_4=\infty$ and write the background metric as
\cite{Giombi:2017cqn}
\begin{eqnarray}
\nonumber   && d s^2_{\text{AdS}_5}=\frac{(1+\frac{|x^i|^2}{4})^2}{(1-\frac{|x^i|^2}{4})^2}d \tau^2+\frac{dx^i d x^i}{(1-\frac{|x^i|^2}{4})^2}\;,\\
\nonumber&& ds_{\text{S}^5}^2= \frac{q^2 dy^a dy^a}{(1+\frac{|y^a|^2}{4})^2}+ \frac{dq^2}{1-q^2}+(1-q^2)d\phi^2 \;, 
\end{eqnarray}
where $\tau$  parameterizes $\mathbb{R}\in$ AdS$_5$ and $y^a\in \mathbb{R}^3$ are the stereographic coordinates of the S$^3$ occupied by the giant graviton. The transverse directions are $x^i$ in $\mathbb{R}^4$ and  $q$, $\phi$ in S$^5$.  The maximal giant graviton solution corresponds to $\phi=i\tau$ and $q=1$. Around this solution, there are small fluctuations \cite{McGreevy:2000cw,Das:2000st}
\begin{equation}
\nonumber    x^i=\delta x^i(\tau,y)\;, q=1+\delta q(\tau,y)\;,\phi=i\tau +\delta \phi (\tau,y)\;,
\end{equation}
which corresponds to fields living on the D3-brane. Further including the variations of the background, the D3-brane action admits the expansion
\begin{equation}
\nonumber    S_{\rm{D3}}=-\frac{N}{2\pi^2}\int d \tau \left(a_0+ L_{\rm{D3}}^{(2, 0)}+ L_{\rm{D3}}^{(0, 1)}+ L_{\rm{D3}}^{(1, 1)}+\ldots \right)\;,
\end{equation}
where $a_0$ is a constant and we have integrated over S$^3$ to obtain a 1d Lagrangian \footnote{More precisely, an orbit average integrating over the $\phi\to\phi+\phi_0$, $\tau\to\tau+\tau_0$ moduli is also needed \cite{Yang:2021kot}.}. In this expansion $ L_{\rm{D3}}^{(m, n)}$ denotes a term with $m$ brane and $n$ bulk fields. Among them, $L_{\rm{D3}}^{(2, 0)}$ is the kinetic term for the defect fields from which we can read off the spectrum. For computing the giant graviton correlators, only three types of defect fields are relevant. They are summarized in Table \ref{tablespectrum}, with conformal dimensions $\widehat{\Delta}$, transverse spins $s$. Under the unbroken R-symmetry, they are in the representation $(r_1,r_1)$  of $SO(4)=SU(2)\times SU(2)$ and carry $SO(2)$ charge $r_2$ \cite{Das:2000st,Imamura:2021ytr}. The bulk spectrum is the same as in \cite{Kim:1985ez} and the relevant fields are collected in Table \ref{tablespectrum}. They couple directly to the brane via the $L_{\rm{D3}}^{(0,1)}$ one-point vertices which have the form
\begin{equation}
\begin{split}
\label{L01}
    L_{\rm{D3}}^{(0,1)}\supset\sum_{k}& \big(c_1 s_k+ c_2 t_{k+4}+c_3 A_{k+1,\tau}\\&+c_4 C_{k+3,\tau}  +c_5 r_{k+2}+c_6 \varphi_{k+2,\tau\tau}\big)\;.
    \end{split}
\end{equation}
Here $c_i$ come from integrating S$^5$ sphere harmonics on S$^3$ and are relevant for R-symmetry.  For the bootstrap calculation, we only need this schematic form and the explicit details of $L_{\rm{D3}}^{(0,1)}$ are left to a companion paper \cite{longpaper}.  It is satisfying to note that this 1d Lagrangian has the form of a probe particle minimally coupled to gauge and gravity fields as one would expect.

\begin{table}[h]
\begin{center}
{\renewcommand{\arraystretch}{1.3}
\begin{tabular}{c c c c || c c c c }
% \hline 
~Bulk~ & $\Delta$ & $~~\ell~~$  & $\,$R rep.$\,$ & ~Defect & ~~$\widehat{\Delta}$~~ &  $~~s~~$  & $\,$ $(r_1,r_2)$ $\,$ \\
 \hline
 \hline
$s_k$ & $k$ &  0  & $[0,k,0]$ & $\chi_{l}$ &$l-1$ & 0 & $(l,-1)$\\
 % \hline
$A_{k,\mu}$ & $k+1$ &  1  & $[1,k-2,1]$ & $x^i_{l}$ &$l$ &1 & $(l-1,0)$ \\ % \hline
$\varphi_{k,\mu\nu}$ & $k+2$ & 2 & $[0,k-2,0]$ &  $\bar{\chi}_{l}$ &$l+1$ & 0 & $(l-2,1)$\\ $C_{k,\mu}$ & $k+3$ &  1  & $[1,k-4,1]$ & & & &\\ $u_{k}$ & $k+4$ &  0  & $[0,k-4,0]$ & & & &\\ $r_{k}$ & $k+2$ &  0  & $[2,k-4,2]$ & & & &
\end{tabular}}
 \caption{Exchange spectra in bulk and defect channels.}
\label{tablespectrum}
\end{center}
\end{table}

The bulk fields couple to the defect fields via $L_{\rm{D3}}^{(1,1)}$. The relevant vertices contain one $s_k$ and have the form 
\begin{equation}
\begin{split}
\label{L11}
    L_{\rm{D3}}^{(1,1)}\supset\sum_{k,l} & \big( d_1 x^i_{l} \partial_i s_k+d_2 \chi_{l} s_k+ d_3\bar{\chi}_{l} s_k  \\&  +d_4 \partial_{\tau}\chi_{l} s_k+ d_5\partial_{\tau}\bar{\chi}_{l} s_k \big)\;,
    \end{split}
\end{equation}
where $d_i$ are similar to $c_i$ in $L_{\rm{D3}}^{(0,1)}$ and will be detailed in \cite{longpaper}. A special feature here is that we can have  single $\partial_\tau$ derivatives thanks to the 1d nature of the AdS defect. In principle, we would also need the $L_{\rm{D3}}^{(0,2)}$ vertices for two $s_k$. However, it is known in similar examples that direct supergravity calculations yield wrong answers \cite{Gimenez-Grau:2023fcy,Zhou:2024ekb}. Fortunately, this information can be fixed by the bootstrap and is not needed as an input.

\vspace{0.5cm}

\noindent{\bf Bootstrap.} At large $N$ the leading $\mathcal{O}(1/N)$ connected defect correlator can be computed in principle as a sum of tree-level Witten diagrams. However, this requires the explicit details of the complicated effective Lagrangian and is practically very cumbersome. A more efficient strategy is to directly bootstrap the correlators using supersymmetry \cite{Rastelli:2016nze,Rastelli:2017udc} (see \cite{Bissi:2022mrs} for a review). For our problem, the algorithm will be highly similar to \cite{Gimenez-Grau:2023fcy,Chen:2023yvw,Zhou:2024ekb}. Therefore, we will be schematic in outlining the strategy and focus on highlighting the differences. 

\begin{figure}[t!]
    \begin{subfigure}[b]{0.24\textwidth}
        \centering
       \includegraphics[width=\textwidth]{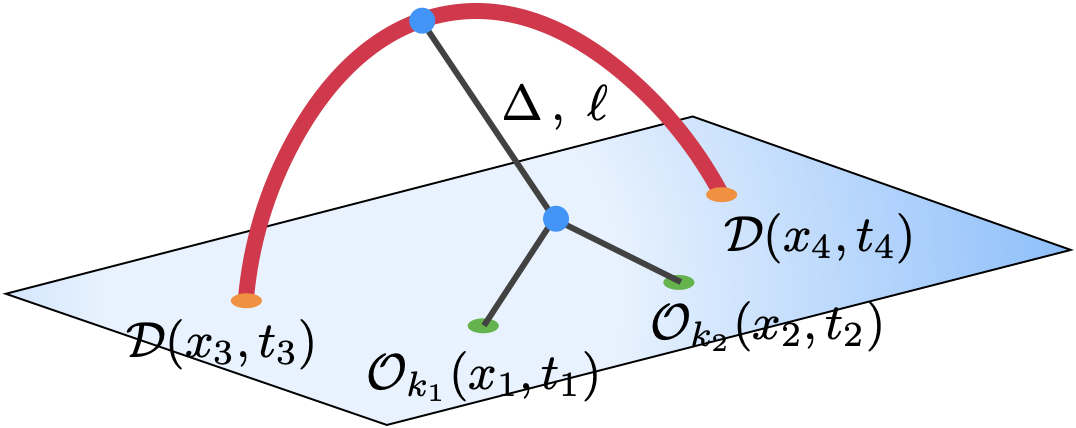}
        \caption{Bulk channel exchange Witten diagrams.}
    \end{subfigure}%
    ~ 
    \begin{subfigure}[b]{0.24\textwidth}
        \centering
       \includegraphics[width=\textwidth]{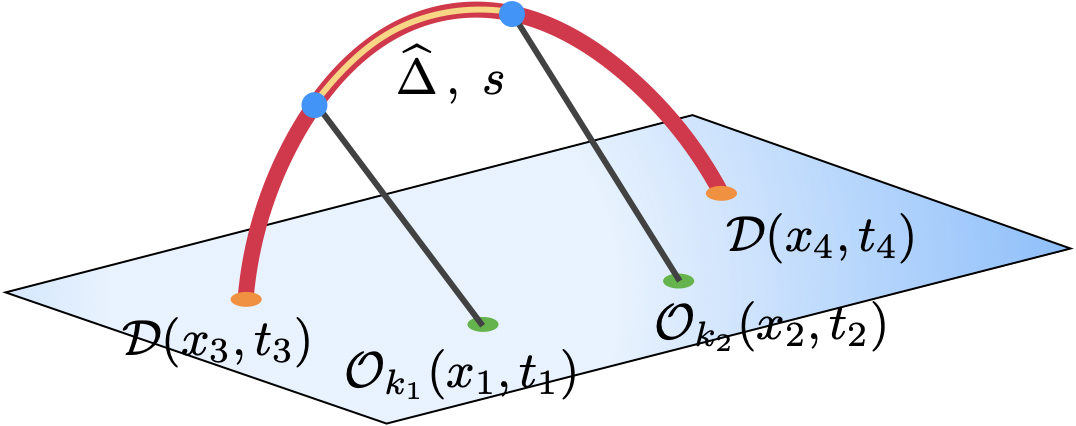}
        \caption{Defect channel exchange Witten diagrams.}
    \end{subfigure}
    ~ 
    \begin{subfigure}[b]{0.24\textwidth}
        \centering
       \includegraphics[width=\textwidth]{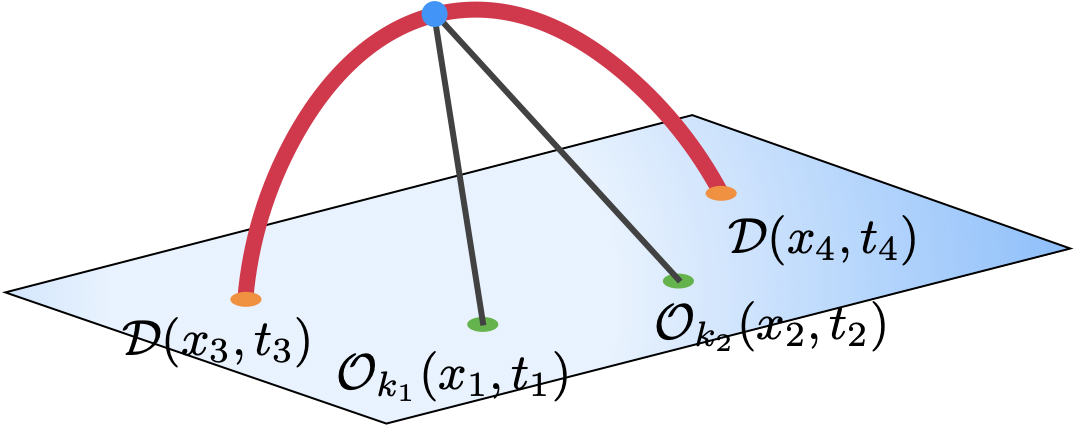}
        \caption{Contact Witten diagrams.}
    \end{subfigure}
    \caption{Tree-level Witten diagrams. The giant graviton travels along the red geodesic line in AdS$_5$ and interacts with supergravitons in three types of processes.}\label{Fig_Wittendiagrams}
\end{figure}

We start with an ansatz
\begin{equation}
    G_{k_1 k_2}^{\rm ansatz}=G^{\rm bulk}_{k_1 k_2}+G^{\rm defect}_{k_1 k_2}+G^{ \rm contact}_{k_1 k_2}\;,
\end{equation}
where the three parts respectively correspond to exchange processes in the bulk and defect channels, and contact interactions (see Figure \ref{Fig_Wittendiagrams}). The exchange parts are sums of possible exchange Witten diagrams multiplied with R-symmetry polynomials capturing the exchanged R-symmetry representations. In the bulk channel, it takes the form 
\begin{equation}
G^{\rm bulk}_{k_1 k_2} =  \sum_{\mathcal{X}\in \mathcal{S}} \mu_{\mathcal{X}} E_{k_1 k_2}^{\Delta_{\mathcal{X}},\ell_{\mathcal{X}}} Q^{k_1k_2}_{\mathcal{X}}\;,
\end{equation}
where $\mathcal{S}$ is chosen from Table \ref{tablespectrum} and $\mu_{\mathcal{X}}$ are unknowns. Using the method of \cite{Rastelli:2017ecj,Gimenez-Grau:2023fcy}, the exchange Witten diagrams can be reduced to finitely many contact Witten diagrams. The latter is known in closed form for any defect dimension $p$ ($p=0$ in our case) \cite{Gimenez-Grau:2023fcy}. The R-symmetry polynomial $Q_{\mathcal{X}}$ can be obtained by solving the $SO(6)$ R-symmetry Casimir equation and are the same as those in \cite{Nirschl:2004pa} for supergraviton four-point functions. Definitions and explicit expressions for Witten diagrams and R-symmetry polynomials can be found in the appendix.

The defect channel is similar and we write it as
\begin{equation}
 \nonumber    G^{\rm defect}_{k_1 k_2} = \!\!\!\!  \sum_{\mathcal{Y}\in \widehat{\mathcal{S}},i=0,1,2}\!\!\!\!\!\! \widehat{\mu}_{\mathcal{Y},i}  \widehat{E}_{k_1 k_2,(i)}^{\Delta_{\mathcal{Y}},0} \widehat{Q}^{k_1k_2}_{\mathcal{Y}}+\! \sum_{\mathcal{Z}\in \widehat{\mathcal{S}}} \widehat{\mu}_{\mathcal{Z}} \widehat{E}_{k_1 k_2}^{\Delta_{\mathcal{Z}},1} \widehat{Q}^{k_1k_2}_{\mathcal{Z}}\;,
\end{equation}
where the defect fields selected from Table \ref{tablespectrum} are split into two groups according to their transverse spins $s$. For $s=0$, we note from (\ref{L11}) that the bulk-defect vertices contain both zero- and one-derivative couplings. Therefore, these defect exchange Witten diagram can have $i=0,1,2$ derivatives. It turns out that the defect exchange Witten diagrams can also be reduced to finitely many contact Witten diagrams. Furthermore, $\widehat{E}_{k_1 k_2}^{\Delta_{\mathcal{Z}},1}$ with $s=1$ are proportional to $\widehat{E}_{k_1+1, k_2+1,(0)}^{\Delta_{\mathcal{Z}},0}$ with $s=0$ \cite{Gimenez-Grau:2023fcy} (see (\ref{Espin1}) in the appendix). As in the bulk channel, the R-symmetry polynomials can be obtained from the Casimir equations of the residual $SO(4)\times SO(2)$ R-symmetry and unknown coefficients $\widehat{\mu}_{\mathcal{Y},i}$ and $\widehat{\mu}_{\mathcal{Z}}$ are introduced.   

A few comments are in order about how $\mathcal{S}$ and $\widehat{\mathcal{S}}$ are determined. The exchanged fields first of all need to be compatible with R-symmetry selection rules. This condition alone already ensures that the lists are finite. Moreover, the exchange vertices must be non-extremal in order to guarantee the finiteness of the effective action \cite{Rastelli:2017udc}. More precisely, this means $\Delta-\ell<k_1+k_2$ for the bulk channel and $\widehat{\Delta}-s<\min\{k_1,k_2\}$ for the defect channel.    

For the remaining contact part, we use the ansatz
\begin{equation}
    G^{ \rm contact}_{k_1 k_2}=  C_{k_1 k_2}^{(0)} \sum_{i,j \geq 0}^{i+j\leq \min\{k_1,k_2\}}\bar\mu_{ij}\sigma^i \tau^j \;,
\end{equation}
where we included all R-symmetry structures and require contact Witten diagrams to have zero derivatives at the vertices. The latter is because higher-derivative contact Witten diagrams would otherwise dominate the high energy limit which is not expected. 

Finally, we impose the condition of superconformal symmetry (\ref{defGandH}). To use it in practice, we set $\bar{\alpha}=1/\bar{z}$ and require the correlator to have no dependence on $\bar{z}$ \footnote{This is the superconformal Ward identity of which (\ref{defGandH}) is the solution.}. We find that this fixes the correlator $G_{k_1k_2}$ up to an overall constant. By using the fact that $\mu_{\mathcal{X}}$ are the products of three-point and one-point function coefficients, we can further reduce these undetermined coefficients to just one and fix it using (\ref{action}). For example, we get for $k_1=k_2=2$
\begin{equation}
    G_{22,{\rm free}}= \frac{2 t_{12}^2 (\tau -\sigma  \tau  U+\sigma  V)}{ N (P_1\cdot \mathbb{N}\cdot P_1)(P_2\cdot \mathbb{N}\cdot P_2) U }\;,
\end{equation}
\begin{equation}
    H_{22}= \frac{2V \left(V^2-2 V \log V-1\right)}{N(P_1\cdot \mathbb{N}\cdot P_1)^2(P_2\cdot \mathbb{N}\cdot P_2)^2U (1-V)^3}\;.
\end{equation}

Our results pass nontrivial tests which we will discuss in detail in \cite{longpaper}. We can compare $G_{k_1k_2,{\rm free}}$ in (\ref{defGandH}) with independent calculations in the free theory \footnote{More rigorously speaking, we should compare the meromorphic twisted correlator after setting $\bar{\alpha}=1/\bar{z}$ which is coupling independent \cite{bllprv13}.}. Free correlators with $k_1=k_2=p$ have been computed in \cite{Jiang:2019xdz} and agree perfectly with our supergravity result \footnote{More precisely, the results for even $p$ differ  by a term which is proportional to the product of two one-point functions. But this is because \cite{Jiang:2019xdz} considered $U(N)$ gauge group instead of $SU(N)$.}. From our results, we can also extract the one-point functions of the bulk fields. These one-point functions can be computed independently from (\ref{L01}) but were not used in our calculations. We find that these two computations are in full agreement. Finally, the integrated correlator for $k_1=k_2=2$ has been computed in \cite{Brown:2024tru} using localization. Our results are consistent with their prediction at strong coupling.

\vspace{0.5cm}

\noindent{\bf Hidden symmetry.} Our bootstrap results encode a remarkable hidden structure which allows us to unify all defect two-point functions into a single generating function. This structure is similar to the 10d hidden conformal symmetry observed in supergraviton four-point functions \cite{Caron-Huot:2018kta}, but extends the latter in nontrivial ways. 

Let us first state our result. We combine $P$ and $t$ into twelve dimensional vectors in the embedding space for $\mathbb{R}^{10}$
\begin{equation}
    Z_i=(P_i,t_i)\;, \label{defZ}
\end{equation}
Then we define the following replacement rules ${\bf T}$
\begin{eqnarray}
       &&P_1\cdot P_2\to Z_1\cdot Z_2\;, \label{repla1}\\
       && P_a\cdot \mathbb{N}\cdot P_b\to Z_a\cdot (\mathbb{N}+\mathbb{M})\cdot Z_b\;,\;\; a,b=1,2\;. \label{repla2}
\end{eqnarray}
 Note that the lowest reduced correlator $H_{22}$ depends only on $x$ and not on $t$. In fact, the defect correlator nature dictates that it only depends on the LHS of the replacement rules. We perform this replacement on $H_{22}$
\begin{equation}
 {\bf H}(x_i;t_i;\lambda_1,\lambda_2)={\bf T}[H_{22}]\big|_{t_1\to\frac{\lambda_1}{\sqrt{2}}t_1,t_2\to\frac{\lambda_2}{\sqrt{2}}t_2}\;,
\end{equation}
where we also rescale $t_{1,2}$ by $\lambda_{1,2}$ to more easily keep track of their dependence. We claim that this promotes $H_{22}$ into a generating function. It turns out that remarkably all correlators with higher weights are obtained from $\bf{H}$ by Taylor expanding in $t_{1,2}$ (practically in $\lambda_{1,2}$)
\begin{equation}
    H_{k_1k_2}=\frac{\sqrt{k_1k_2}}{2}{\bf H}(x_i;t_i;\lambda_1,\lambda_2)\big|_{\lambda_1^{k_1-2}\lambda_2^{k_2-2}}\;,
\end{equation}
where $(\ldots)\big|_{\lambda_1^a\lambda_2^b}$ means to take the coefficient of $\lambda_1^a\lambda_2^b$. 

Our observation bears significant resemblance with that of \cite{Caron-Huot:2018kta} but is different in a few important details. While in both cases (\ref{defZ}) is used to put $P$ and $t$ on the same footing, only a single replacement rule (\ref{repla1}) exists in \cite{Caron-Huot:2018kta} and is applied to all pairs of points $(i,j)$ to obtain a generating function. Further expanding in all $t_i$ would then lead to increase of quantum numbers also at 3 and 4, which is not wanted here. In our case, we instead leave individual $P_{3,4}$ and $t_{3,4}$ untouched and uplift the effect of giant gravitons by promoting the 4d spacetime projector $\mathbb{N}$ into a 10d projector $\mathbb{N}+\mathbb{M}$. This leads to a defect generalization of the hidden symmetry found in \cite{Caron-Huot:2018kta}. Our prescription is also compatible with the intuitive but imprecise picture that a Weyl transformation maps a propagating giant graviton in AdS$_5\times$S$^5$ to an $\mathbb{R}^4$ defect inside $\mathbb{R}^{10}$ (away from the conformal boundary). Note that the part of the S$^5$ orthogonal to $\mathbb{M}$ is the S$^3$ which becomes a part of the 4d defect. This explains the asymmetric definitions which we used in (\ref{defN}) and (\ref{defM}).

\vspace{0.5cm}

\noindent{\bf Weak coupling.} A fascinating feature of the strong coupling hidden symmetry observed in \cite{Caron-Huot:2018kta} is that it also holds at weak coupling, albeit in a slightly different manner. At each loop order, perturbative corrections to the supergraviton four-point function can be expressed as integrals using Lagrangian insertion method. It was shown in \cite{Caron-Huot:2021usw} that hidden symmetry promotes the integrands of $k_i=2$ into  generating functions for integrands of all other four-point correlators \footnote{The same was also observed for 4d $\mathcal{N}=2$ SCFTs. An 8d hidden conformal symmetry was first found at strong coupling in \cite{Alday:2021odx} and then identified at weak coupling for one-loop corrections in \cite{Du:2024xbd}.}. We find a similar phenomenon at weak coupling for giant graviton correlators after extending the replacement rules {\bf T}. 

Let us expand (\ref{defGandH}) in coupling and denote the $\ell$-th loop correction as $H_{k_1k_2}^{(\ell)}$. The Lagrangian insertion method expresses it as an $\ell$-fold conformal integral which has integrand $I_{k_1k_2}^{(\ell)}$ and is integrated over  insertion points $x_{p=5,\ldots,4+\ell}$. For example, the one-loop integrand is \cite{Jiang:2019xdz}
\begin{equation}
    I_{22}^{(1)}=-\frac{x_{34}^6}{4\pi^2x_{12}^2x_{13}^2x_{14}^2x_{23}^2x_{24}^2}\times \frac{1}{x_{15}^2x_{25}^2x_{35}^2x_{45}^2}\;.
\end{equation}
To uplift $I_{22}^{(\ell)}$, we  introduce $Z_p=(P_p,0)$ also for the insertion points. The extended replacement rules, which we denote as ${\bf T'}$,  are obtained by enlarging the allowed labels values in (\ref{repla2}) to the set $\{1,2,5,\ldots,4+\ell\}$  in which  all points but 3 and 4 are included. Clearly, $I_{22}^{(\ell)}$ is an $(\ell+2)$-point defect correlator where each point has dimensions 4 and is R-symmetry neutral. Therefore, the variables at the LHS of the enlarged replacement rules are sufficient to uniquely express these integrands. It is also convenient to note that the lack of R-symmetry components in $Z_p$ of the insertion points renders the replacement for $P_m\cdot \mathbb{N}\cdot P_n$  trivial unless both $m$ and $n$ take values 1 or 2. Finally, we conjecture that at any loop the integrand  $I_{k_1k_2}^{(\ell)}$ is given by
\begin{equation}
\begin{split}
 {\bf I}^{(\ell)}={}&{\bf T'}[I_{22}^{(\ell)}]\big|_{t_1\to\frac{\lambda_1}{\sqrt{2}}t_1,t_2\to\frac{\lambda_2}{\sqrt{2}}t_2}\;,\\
I_{k_1k_2}^{(\ell)}={}&\frac{\sqrt{k_1k_2}}{2}{\bf I}^{(\ell)}\big|_{\lambda_1^{k_1-2}\lambda_2^{k_2-2}}\;.
\end{split}
\end{equation}

We find nontrivial evidence supporting our conjecture. For $\ell=1$, the explicit generating function reads
\begin{equation}
\nonumber    {\bf I}^{(1)}=-\frac{(\frac{x_{13}^2x_{14}^2}{x_{34}^2}+\frac{\lambda_1^2t_{13}t_{14}}{t_{34}})^{-1}(\frac{x_{23}^2x_{24}^2}{x_{34}^2}+\frac{\lambda_2^2t_{23}t_{24}}{t_{34}})^{-1}x_{34}^2}{4\pi^2(x_{12}^2-\lambda_1\lambda_2 t_{12})x_{15}^2x_{25}^2x_{35}^2x_{45}^2}\;.
\end{equation}
Expanding in $\lambda_{1,2}$ and extrcting the $\lambda_1^{p-2}\lambda_2^{p-2}$ coefficient, we reproduce exactly $I_{pp}^{(1)}$ computed in \cite{Jiang:2019xdz}. For $\ell=2$, we write down a generating function by uplifting the known $I_{22}^{(2)}$ \cite{Jiang:2019xdz}. The integrand $I_{pp}^{(2)}$ predicted by hidden symmetry also agrees with independent Feynman diagram calculation \cite{twoloopDDpp}. It would be interesting to test our conjecture for more general weights and also at higher loops.

\vspace{0.5cm}

\noindent{\bf Discussions and outlook.} In this paper, we used bootstrap techniques to determine all four-point functions of light supergravitons with two maximal giant gravitons. A key insight here is to view the heavy giant gravitons as a zero dimensional defect. Our results lead to a number of research avenues for future explorations. 

An immediate generalization is to apply the defect insight to correlators involving non-maximal giant gravitons \cite{Chen:2019gsb}. It would be interesting to see how the results depend on the size of the giant gravitons and if the observed hidden structure persists. By performing conformal block decompositions, we can extract from the giant graviton correlators rich CFT data (such as anomalous dimensions and OPE coefficients) at strong coupling. In this work we considered four-point functions corresponding to AdS amplitudes of a D3 brane emitting two supergravity modes. A natural extension is to examine four-point functions of four giant gravitons \cite{Corley:2001zk,Corley:2002mj,Bhattacharyya:2008xy,Vescovi:2021fjf,Shahpo:2021xax}, which describe the interaction between two D3 branes mediated by the exchange of supergravity modes. Such configurations are expected to exhibit interesting phase transitions as the operator locations are varied. We believe that analytic bootstrap techniques together with the defect intuition will continue to be powerful enough to constrain such four-point functions at strong coupling.

There are also many other extensions which can be studied by exploiting technologies readily available in the literature. Starting from our tree-level supergravity result, we can further use unitarity method to construct loop corrections in AdS \cite{Aharony:2016dwx,Chen:2024orp}. We can also study stringy corrections using localization constraints \cite{Brown:2024tru} and flat-space amplitudes which are related by the flat-space limit formula \cite{Alday:2024srr}. Another interesting direction involves  correlators with two giant gravitons and additional
light operators. These are natural analogues of higher-point correlators in defect-free CFTs considered in \cite{Goncalves:2019znr,Alday:2022lkk,Goncalves:2023oyx,Alday:2023kfm,Cao:2023cwa,Cao:2024bky,Huang:2024dxr,Goncalves:2025jcg}, of which the strategies can be adapted. Gaint graviton correlators can also be similarly considered in other maximally supersymmetric holographic setups, particularly in AdS$_4$/CFT$_3$ \cite{Chen:2019kgc,Yang:2021hrl} and AdS$_3$/CFT$_2$ correspondences.

Finally, another particularly interesting intersection is with integrabililty and concerns the OPE coefficients of two giant gravitons and a non-BPS operator. Integrability describes giant gravitons as the initial or boundary states (therefore also indicating a defect picture). These states play an important role in non-equilibrium statistical mechanics and serve as initial states for integrable systems in quantum quench protocols. The OPE coefficients correspond to the overlap between an integrable boundary state and an energy eigenstate (associated with the non-BPS operator), and is known as the exact $g$-function for excited states. The latter plays a central role in characterizing quantum many-body systems with impurities, such as in the Kondo problem, and its computation is highly nontrivial, especially for theories with non-diagonal scattering. Current integrability-based approaches, such as the TBA method \cite{Jiang:2019xdz,Jiang:2019zig}, face both technical challenges and a lack of reliable data for validation. Recently, lattice-based methods for computing the $g$-function have been proposed \cite{He:2024asy}, offering a partial resolution to these difficulties. With the results of this work, we can now provide invaluable benchmark data to test integrability predictions.

\vspace{0.5cm}

\vspace{0.5cm} 
\noindent{\bf Acknowledgements.} The work of J.C. and X.Z. is supported by the NSFC Grant No. 12275273, funds from Chinese Academy of Sciences, University of Chinese Academy of Sciences, and the Kavli Institute for Theoretical Sciences. The work of X.Z. is also supported by the Xiaomi Foundation. The work of Y.J. is partly supported by Startup Funding no. 4007022326 of Southeast University. We thank the hospitality of Institute of Modern Physics at Northwest University where part of the work was done. This work is also supported by the NSFC Grant No. 12247103.

\appendix

\section{Witten diagrams}
We collect explicit results for Witten diagrams needed for performing the bootstrap analysis and leave a more detailed account to \cite{longpaper}. To simplify the calculation, the two giant graviton operators are inserted at $x_3=0$ and $x_4=\infty$ which are interpolated in the bulk by the geodesic line
\begin{equation}
    x^{1,2,3,4}(\tau)=0\;,\quad x^0(\tau)=e^\tau\;.
\end{equation}
The bulk-to-boundary propagator from a point on the boundary $x_i$ to the geodesic line is
\begin{equation}
\nonumber    K_{\Delta_i}^{B\partial}(x_i,x_0(\tau))=\left(\frac{x_0}{x_0^2+x_i^2}\right)^{\Delta_i}\;.
\end{equation}

We first consider contact Witten diagrams with $n$ derivatives in the vertex
\begin{equation}
\nonumber    C_{\Delta_1 \Delta_2}^{(n)}=\int d \tau K_{\Delta_1}^{B\partial}(x_1,x_0(\tau))(\partial_{\tau})^n K_{\Delta_2}^{B\partial}(x_2,x_0(\tau))\;.
\end{equation}
It is convenient to factor out the one-point functions so that we can work with functions of cross ratios which are denoted using calligraphic letters
\begin{equation}
    C_{\Delta_1 \Delta_2}^{(n)}=\frac{\mathcal{C}_{\Delta_1 \Delta_2}^{(n)}(U,V)}{\prod_{i=1}^2 (-P_i\cdot \mathbb{N}\cdot P_i)^\frac{\Delta_i}{2}}\;.
\end{equation}
It is straightforward to show that the contact Witten diagrams with $n=0$ coincide with the ones considered in \cite{Gimenez-Grau:2023fcy} after setting the defect dimension $p=0$
\begin{equation}
\begin{aligned}
    \mathcal{C}_{\Delta_1 \Delta_2}^{(0)}&=\frac{\pi^{\frac{1}{2}}\Gamma( \frac{\Delta_1+\Delta_2}{2})}{2^{\Delta_1+\Delta_2}\Gamma( \frac{\Delta_1+\Delta_2+1}{2})} \\&\times {}_2 F_1 \left(\Delta_1,\Delta_2,\frac{\Delta_1+\Delta_2+1}{2},-\frac{(1-\sqrt{V})^2}{4\sqrt{V}} \right)\;.
    \end{aligned}
\end{equation}
One can also show that the contact Witten diagrams with $n>0$ can be reduced to the $n=0$ case
\begin{equation}
\begin{split}
\nonumber    \mathcal{C}_{\Delta_1 \Delta_2}^{(1)}={}&\Delta_1 \left(\frac{2\sqrt{V}}{1-V}  \mathcal{C}_{\Delta_1+1,\Delta_2-1}^{(0)}-\frac{1+V}{1-V} \mathcal{C}_{\Delta_1 \Delta_2}^{(0)}\right)\;,\\
     \mathcal{C}_{\Delta_1 \Delta_2}^{(2)}={}&\Delta_1 \Delta_2 \left(\frac{2(1+V)}{\sqrt{V}}  \mathcal{C}_{\Delta_1+1,\Delta_2+1}^{(0)}-\mathcal{C}_{\Delta_1 \Delta_2}^{(0)}\right)\;.
\end{split}
\end{equation}

By extending the method of \cite{Rastelli:2017ecj,Gimenez-Grau:2023fcy}, exchange Witten diagrams can be expressed as finitely many contact Witten diagrams. Let us first consider the bulk channel and exchange a field with spin $\ell$
\begin{equation}\label{EDeltaell}
   E_{\Delta_1\Delta_2}^{\Delta,\ell}=\int d \tau I^{\mu_1\ldots\mu_{\ell}}_{\Delta}(x_1,x_2,\tau) \delta_{\mu_1,\tau} \cdots \delta_{\mu_\ell,\tau}\;.
\end{equation}
Here we focus on the integral on the line and $I^{\mu_1\ldots\mu_{\ell}}_{\Delta}$ is the three-point function part obtained from integrating out the bulk point. When $\Delta_1+\Delta_2-\Delta+\ell\in 2\mathbb{Z}_+$, the function  $I^{\mu_1\ldots\mu_{\ell}}_{\Delta}$ becomes a finite sum of products of bulk-to-boundary propagators, which reduces (\ref{EDeltaell}) to contact Witten diagrams. Explicitly, for $\ell=0,1$ we have
\begin{equation}
\nonumber
    \mathcal{E}_{\Delta_1 \Delta_2}^{\Delta_1+\Delta_2+\ell-2m,\ell}=\sum_{i=0}^{m-1} \frac{U^{i-m}}{V^{\frac{i-m}{2}}}  a_i^{(\ell)} \mathcal{C}^{(\ell)}_{\Delta_1+i-m,\Delta_2+i-m}\,,
\end{equation}
where $m\in\mathbb{Z}_+$ and
\begin{equation}
    \begin{aligned}
 a_{i}^{(\ell)} &=  \frac{\Gamma (m)\Gamma \left(i-m+\Delta _1\right) \Gamma \left(i-m+\Delta _2\right)}{2^{2-\ell} \Gamma (\Delta _1) \Gamma (\Delta
   _2) \Gamma (i+1)}\\& \quad \quad \times  \frac{ \Gamma \left(\Delta _1+\Delta _2-m+\ell-2\right) }{\Gamma \left(\Delta _1+\Delta _2+i-2 m+\ell-1\right)}\;.
\end{aligned}
\end{equation}
The case of $\ell=2$ is more involved and here we only present the simplest examples
\begin{equation}
\begin{aligned}
\nonumber
     &\mathcal{E}_{\Delta_1 \Delta_2}^{\Delta_1+\Delta_2,2}=\frac{-2 \Delta_1 \Delta_2 (1-V)^2}{(\Delta_1+\Delta_2)(\Delta_1+\Delta_2-1)U\sqrt{V}}\mathcal{C}_{\Delta_1+1,\Delta_2+1}^{(0)}\;,\\& \mathcal{E}_{\Delta_1 \Delta_2}^{\Delta_1+\Delta_2-2,2}=\frac{-2}{(\Delta_1+\Delta_2-2)U}\bigg(\frac{(1-V)^2}{U} \mathcal{C}_{\Delta_1\Delta_2}^{(0)}       \\& \quad + \frac{(1-V)^2}{\sqrt{V}}\frac{(\Delta_1 \Delta_2-2(\Delta_1+\Delta_2)+3)}{\Delta_1+\Delta_2-6} \mathcal{C}_{\Delta_1+1,\Delta_2+1}^{(0)}   \\&\quad +
     \bigg(\frac{3-2\Delta_1+\Delta_2}{\Delta_1+\Delta_2-6}\sqrt{V}\mathcal{C}_{\Delta_1-1,\Delta_2+1}^{(0)}  +(\Delta_1\leftrightarrow \Delta_2)\bigg)     \bigg)\;.
     \end{aligned}
\end{equation}
In the defect channel, we first consider the case with zero transverse spins
\begin{equation}
\begin{aligned}
\nonumber
    \widehat{E}_{\Delta_1\Delta_2,(n)}^{\widehat{\Delta},0}=\int d \tau d\tau' & K_{\Delta_1}^{B\partial}(x_1,x_0(\tau))  K_{\Delta_2}^{B\partial}(x_2,y_0(\tau')) \\&\times \mathbb{D}_{(n)}K^{BB}_{\widehat{\Delta}}(x_0(\tau),y_0(\tau'))\;.
    \end{aligned}
\end{equation}
Here $n$ is the number of derivatives and 
\begin{equation}
\nonumber    \mathbb{D}_{(0)}=1\;,\quad \mathbb{D}_{(1)}=\partial_{\tau}\;,\quad \mathbb{D}_{(2)}=-\partial_{\tau}\partial_{\tau'}\;.
\end{equation}
The bulk-to-bulk propagator on the geodesic line is
\begin{equation}
\nonumber K^{BB}_{\widehat{\Delta}}(x_0,y_0)=\frac{1}{2\widehat{\Delta}} u^{-\widehat{\Delta}} {}_2 F_1 \left( \widehat{\Delta},\widehat{\Delta}+\frac{1}{2};2\widehat{\Delta}+1,-\frac{4}{u}\right)\;,
\end{equation}
where $u=\frac{(x_0-y_0)^2}{x_0 y_0}$. When $\Delta_1-\widehat{\Delta}\in 2\mathbb{Z}_+$, the defect channel exchange Witten diagrams reduce to a finite sum of contact Witten diagrams
\begin{equation}
     \widehat{\mathcal{E}}_{\Delta_1\Delta_2,(n)}^{\Delta_1-2m,0}=\sum_{i=0}^{m-1} \widehat{a}_i \mathcal{C}_{\Delta_1+2i-2m,\Delta_2}^{(n)}\;,
\end{equation}
where
\begin{equation}
\nonumber
    \widehat{a}_i=-\frac{(-1)^i}{4 i!} \Gamma (m) \left(\Delta _1\right)_{-m} \left(-2 i+2 m-\Delta _1+1\right)_{i-1}\;.
\end{equation}
Similarly, the diagram also truncates when $\Delta_2-\widehat{\Delta}\in 2\mathbb{Z}_+$. A minor subtlety is that for $n=0$ and $\widehat{\Delta}=0$, the defect exchange Witten diagram needs to be regularized by multiplying with $\widehat{\Delta}$ and then taking the limit $\widehat{\Delta}\to 0$. This gives the correct answer which is a product of two one-point functions. Finally, exchange diagrams with nonzero transverse spins are related to the $s=0$ case by shifting external dimensions and multiplying simple kinematic factors of cross ratios
\cite{Gimenez-Grau:2023fcy}
\begin{equation}\label{Espin1}
    \widehat{\mathcal{E}}_{\Delta_1\Delta_2}^{\widehat{\Delta},1} = 2 \Delta_1 \Delta_2 \frac{1-U+V}{\sqrt{V}}  \widehat{\mathcal{E}}_{\Delta_1+1,\Delta_2+1,(0)}^{\widehat{\Delta},0}\;.
\end{equation}

\section{R-symmetry polynomials}

To discuss R-symmetry polynomials, it is convenient to also extract a kinematic factor so that the result can be expressed in terms of cross ratios
\begin{equation}
\begin{split}
\nonumber
    Q_{\mathcal{X}}^{k_1k_2}={}&\prod_{i=1}^2\bigg(\!- \frac{ t_i \cdot \mathbb{M}\cdot t_i}{2}\bigg)^{\frac{k_i}{2}}  \mathcal{Q}_{\mathcal{X}}^{k_1 k_2}(\sigma,\tau)\;,\\
    \widehat{Q}_{\{\mathcal{Y},\mathcal{Z}\}}^{k_1k_2}={}&\prod_{i=1}^2\bigg(\!- \frac{ t_i \cdot \mathbb{M}\cdot t_i}{2}\bigg)^{\frac{k_i}{2}}  \widehat{\mathcal{Q}}_{\{\mathcal{Y},\mathcal{Z}\}}(\sigma,\tau)\;.
\end{split}
\end{equation}
Here $\mathcal{Q}$ and $\widehat{\mathcal{Q}}$ are the R-symmetry polynomials in the bulk and defect channel respectively.

As we have mentioned in the main text, the bulk channel R-symmetry polynomials coincide with those of four light supergravitons in \cite{Alday:2020dtb}. This can be seen from the fact that they satisfy the same Casimir equation and obey the same boundary condition dictated by the exchange of representations in the 12 channel. In particular, the large R-charges of 3 and 4 do not matter because the Casimir equation only depends on the difference. 

In the defect channel, the R-symmetry polynomials need to satisfy two Casimir equations which come from the $SO(4)$ and $SO(2)$ factors of the unbroken R-symmetry respectively. These two equations can be disentangled by using a particular combination of R-symmetry cross ratios and the separation of variables gives us two decoupled ODEs (details will be given in \cite{longpaper}). Solving these equations with appropriate boundary conditions, we find the defect channel R-symmetry polynomials are
\begin{equation}
    \widehat{\mathcal{Q}}_{r_1,r_2}=\left(\frac{\sigma}{\tau} \right)^{\frac{r_2}{2}} (-1)^{r_1} C_{r_1}^{1}\left(\frac{\sigma+\tau-1}{2\sqrt{\sigma \tau}}\right)\;.
\end{equation}
Here $C^a_b$ is the Gegenbaur polynomial and we have relabelled the polynomials with the $SU(2)$ spins $r_1$ and $SO(2)$ charges $r_2$. In fact, the solution has a two-fold degeneracy related by $\sigma\leftrightarrow\tau$ which corresponds to $3\leftrightarrow 4$ interchange. The defect channel exchange Witten diagrams have definite parities under $3\leftrightarrow 4$. This requires the corresponding R-symmetry polynomials to take symmetric (or anti-symmetric) combinations so that the product is invariant under $3$ and $4$ interchange.

\bibliography{refs} 
\bibliographystyle{utphys}
\end{document}